# Noncommutative metasurfaces enabled diverse quantum path entanglement of structured photons

Yan Wang, Yichang Shou, Jiawei Liu, Qiang Yang, Shizhen Chen, Weixing Shu, Shuangchun Wen, Hailu Luo*

Laboratory for Spin Photonics, School of Physics and Electronics, Hunan University, Changsha 410082, China

*hailuluo@hnu.edu.cn

## Abstract

Quantum entanglement, a fundamental concept in quantum mechanics, lies at the heart of many current and future quantum technologies. A pivotal task is generation and control of diverse quantum entangled states in a more compact and flexible manner. Here, we introduce an approach to achieve diverse path entanglement by exploiting the interaction between noncommutative metasurfaces and entangled photons. Different from other path entanglement, our quantum path entanglement is evolvement path entanglement of photons on Poincaré sphere. Due to quantum entanglement between idler photons and structured signal photons, evolvement path of idler photons on the fundamental Poincaré sphere can be nonlocally mirrored by structured signal photons on any high-order Poincaré sphere, resulting in quantum path entanglement. Benefiting from noncommutative metasurfaces, diverse quantum path entanglement can be switched across different higher-order Poincaré spheres using distinct combination sequences of metasurfaces. Our method allows for the tuning of diverse quantum path entanglement across a broad spectrum of quantum states, offering a significant advancement in the manipulation of quantum entanglement.

## Introduction

Quantum entanglement is a fundamental resource in quantum information science [1-3], enabling a range of applications from secure communication to advanced computation [4-7]. Benefiting from different degree of freedom (DOF) of photon, quantum entanglement has been realized in several DOFs, including polarization [8], orbit angular momentum (OAM) [9,10], path [11,12], and frequency [13]. To extends beyond the traditional qubit framework, recent advancements in quantum optics have highlighted the potential of using multiple DOFs [14,15], such as polarization-OAM [16], path-OAM [17], polarization-path [18]. These approaches have demonstrated promising results, yet it is significant to facilitate the generation and control of diverse quantum entangled states by modulating photons across multiple DOFs in a more compact and flexible manner.

The capacity to precisely manipulate light across multiple DOFs is crucial for quantum entanglement [19,20]. Many methods have been provided to structure light in multi-DOFs [21-28], and metasurfaces have recently emerged as promising candidates for manipulating light at the nanoscale [29-34]. They enable precise manipulation and control over the amplitude, phase, and polarization of light, allowing for the realization of complex optical functions using ultrathin devices [35-42]. The ability of metasurfaces to simultaneously perform multi-DOFs manipulation makes them particularly advantageous for quantum applications, as they can be tailored to generate specific quantum states with high fidelity [43-53]. However, the full potential of metasurfaces in the context of quantum entanglement can be further explored.

Here, we experimentally demonstrate noncommutative metasurfaces can lead to different optical responses to realize the flexible manipulation of diverse quantum path entanglement. By the interaction between noncommutative metasurfaces and entangled photons, signal photons can be tailored into any structured photons represented by distinct high-order Poincaré spheres (HOPS) [54]. Any evolvement path of idler photons on the fundamental Poincaré sphere (PS) allows for the non-local mirroring by the structured signal photons on HOPS. This achievement marks a significant step towards the realization of quantum path entanglement using ultrathin device. Due to its non-commutativity, the sequence of metasurfaces is adjusted to switch the diverse path entanglement between different HOPS, which is crucial for practical quantum communication and computation applications. Our approach not only overcomes the limitations of traditional bulk optical elements but also provides a platform for the flexible control of entangled states.

**Results**

**Principle**

Firstly, entangled photon pair can be created from a spontaneous parametric down-conversion (SPDC), and spatially separated into idler photons and signal photons. Their entangled state can be expressed by

$$|\Psi\rangle = \frac{1}{\sqrt{2}}(|H\rangle_i |H\rangle_s + |V\rangle_i |V\rangle_s), \tag{1}$$

where subscript $i$ and $s$ represent idler photons and signal photons, respectively. $H$ and $V$ are horizontal polarization and vertical polarization, respectively. The polarization photons can be represented by fundamental PS [blue sphere in Fig.1], and expressed by

$$|\Phi(\alpha,\beta)\rangle=\cos\frac{\alpha}{2}|L_0\rangle e^{+i\frac{\beta}{2}} + \sin\frac{\alpha}{2}|R_0\rangle e^{-i\frac{\beta}{2}}, \tag{2}$$

$$|H\rangle=|\phi(\alpha,\beta)\rangle+|\phi(\alpha',\beta')\rangle, \tag{3}$$

$$|V\rangle=|\phi(\alpha,\beta)\rangle-|\phi(\alpha',\beta')\rangle, \tag{4}$$

where, $|L_0\rangle$ and $|R_0\rangle$ are left-hand circular polarization and right-hand circular polarization, respectively. $\langle\phi(\alpha',\beta')|\phi(\alpha,\beta)\rangle = 0$, $(\alpha,\beta)$ and $(\alpha',\beta')$ are the azimuthal and polar angles of PS. Substitute Eq. (3) & (4) into Eq. (1), the entangled state can be transformed into

$$|\Psi\rangle = \tfrac{1}{\sqrt{2}}(|\phi(\alpha,\beta)\rangle_i |\phi(\alpha,\beta)\rangle_s + |\phi(\alpha',\beta')\rangle_i |\phi(\alpha',\beta')\rangle_s), \tag{5}$$

here, $|\phi(\alpha,\beta)\rangle$ and $|\phi(\alpha',\beta')\rangle$ represent any two orthogonal polarization states on fundamental PS.

Metasurface, as an artificially two-dimensional material composed of subwavelength structure, can manipulate and control DOF of input electromagnetic field to structure photons with different spatial and polarization features. To increase the number of DOFs that can be manipulated, here, the noncommutativity of multiple metasurfaces is considered. Noncommutative metasurfaces indicate different cascaded sequences of metasurfaces with distinct functions can lead to different optical responses (see details in Supplementary Material). As an example, two metasurfaces, metasurafce $M_A$ is with uniform optical axis distribution with π phase retardation and metasurface $M_B$ designed as q-plate with π phase retardation. But they are not restricted to this particular case and can be adapted to arbitrary metasurfaces with different functionality. When entangled photons in the combination of cascaded metasurfaces with interesting characteristics in the modulation of phase and polarization of photon, the noncommutativity of two metasurfaces will be analyzed (see details in Supplementary Material).

When consider signal photons passes through $M_A$ followed by $M_B$ [Fig.1(e)], the structured signal photons state can be written as

$$|\Psi(\alpha,\beta)\rangle_s=\cos\frac{\alpha}{2}|L_1\rangle e^{+i\frac{\beta}{2}} + \sin\frac{\alpha}{2}|R_1\rangle e^{-i\frac{\beta}{2}}, \tag{6}$$

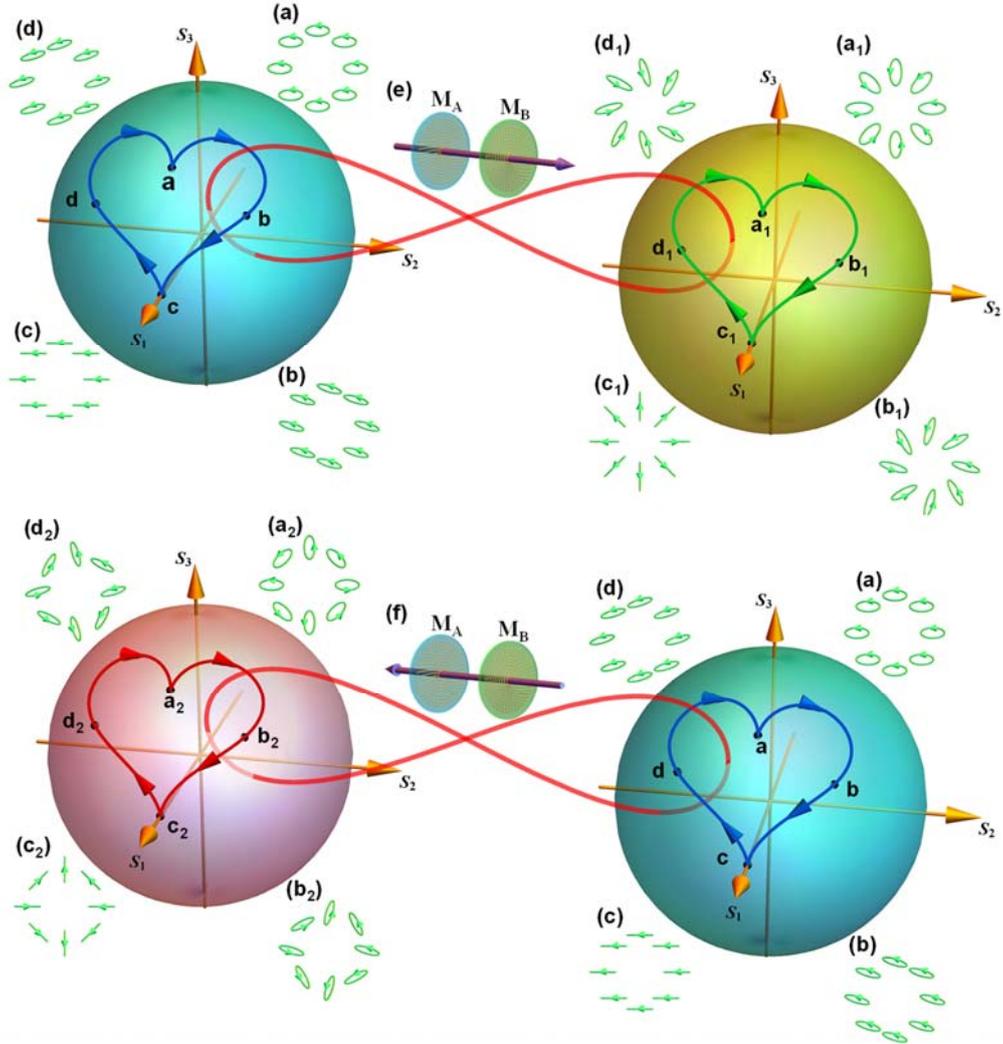

**Fig.1. Noncommutative metasurfaces enabled diverse quantum path entanglement of structured photons.** (a)-(d) The polarization patterns at four specific points along a cardioid path on the fundamental PS (blue), representing the idler photon's homogeneous polarization state across the beam profile. (a1)-(d1) The polarization patterns of structured signal photons at four specific points along a cardioid path on HOPS with order $m = 1$ (green), corresponding to the sequence $M_A M_B$ in (e). (a2)-(d2) The polarization patterns of structured signal photons at four specific points along a cardioid path on HOPS with order $m = -1$ (red), corresponding to the sequence $M_B M_A$ in (f). Cardioid path entanglement can be created and switched across different HOPS by noncommutative metasurfaces.

which $|L_1\rangle = (H + iV)\exp(+im\gamma)/\sqrt{2}$, $|R_1\rangle = (H - iV)\exp(-im\gamma)/\sqrt{2}$, $m = +2q$, and $\gamma$ is the angle difference of optical axis of two metasurfaces with respect to the x axis. If $q = 1/2$, the polarization state of structured signal photons can be represented by the HOPS with order

$m = 1$ [green sphere in Fig.1]. The structured signal photons state and idler photons are mutually entangled

$$|\Psi_1\rangle = \frac{1}{\sqrt{2}}(|\phi(\alpha,\beta)\rangle_i |\Psi(\alpha,\beta)\rangle_s + |\phi(\alpha',\beta')\rangle_i |\Psi(\alpha',\beta')\rangle_s), \qquad (7)$$

where $|\Psi(\alpha,\beta)\rangle_s$ and $|\Psi(\alpha',\beta')\rangle_s$ are any two orthogonal entangled states of structured signal photons on HOPS. When idler photons evolve into any paths on fundamental PS, the corresponding any paths will be mirrored by structured signal photons on HOPS with order $m = 1$, resulting in quantum path entanglement

$$|\Psi_2\rangle = \frac{1}{\sqrt{2}}(|p_1\rangle_i |p_1\rangle_s + |p_2\rangle_i |p_2\rangle_s), \qquad (8)$$

which $|p_1\rangle_i$ and $|p_2\rangle_i$ represent two orthogonal evolvement paths of idler photons on fundamental PS, $|p_1\rangle_s$ and $|p_2\rangle_s$ represent two corresponding orthogonal evolvement paths of structured signal photons on HOPS. As an example, entangled cardioid path between blue sphere and green sphere is shown in Fig.1. In addition, the shape of quantum path entanglement can be various and measured by collapsing idler photons in continuous moment.

When it is considered to change the cascaded order of two metasurfaces into $M_B$ placed before $M_A$ [Fig.1(f)], $m = -2q$. The polarization state of structured signal photons can be represented by the HOPS of $m = -1$ [red sphere in Fig.1]. Analogously, quantum path entanglement is switched from HOPS with order $m = +1$ to $m = -1$, leading to diverse quantum path entanglement, as shown entangled cardioid path between blue sphere and red sphere in Fig.1. Due to diverse quantum path entanglement, each point along the entangled path can be regarded as a distinct basis state between two photons, with all points being mutually entangled. Figures 1(a)-1(d) show polarization distribution of idler photons at specific four points along entangled cardioid path. The polarization patterns of structured signal photons with $m = 1$ and $m = -1$ at specific four points along entangled cardioid path is provided in Figs.1(a1)-1(d1) and Figs.1(a2)-1(d2). It is theoretically demonstrated that noncommutativity of metasurfaces can flexibly create and switch diverse path entanglement state across different HOPS.

# Experimental Setup

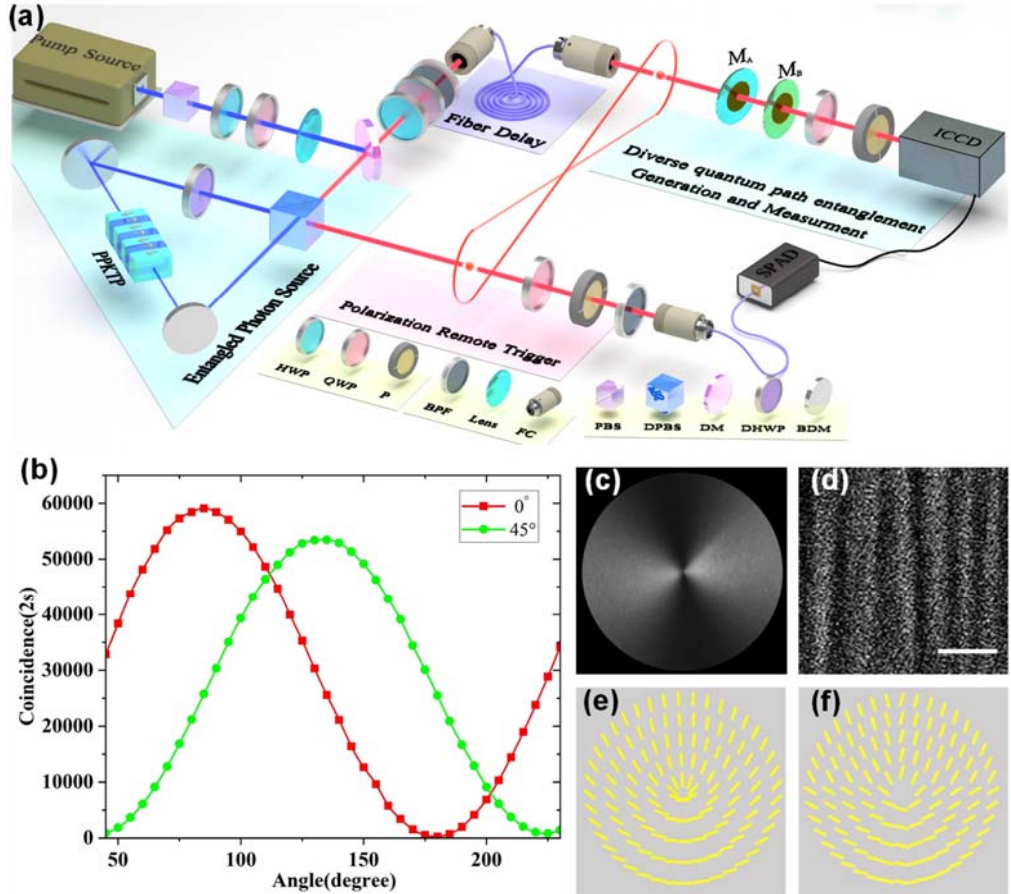

**Fig.2. Experimental apparatus for the generation and measurement of diverse quantum path entanglement of structured photons.** (a) The optical setup includes three stages: entangled photon source, polarization remote trigger, generation and measurement of diverse quantum path entanglement. Polarization entanglement between two photons is created in the type-II down-conversion periodically poled KTiOPO$_4$ (PPKTP) crystal pumped by a single-frequency diode laser at 405 nm. PBS, polarization beam splitter; HWP, half wave plate; QWP, quarter wave plate; DM, dichromatic mirror; DPBS, dual-wavelength polarization beam splitter; DHWP, dual-wavelength half wave plate; BDM, broadband dielectric mirror; BPF, band-pass filter; FC, fiber coupler; P, polarizer; M$_A$, M$_B$, metasurface; SPAD, single photon avalanche detector; ICCD, intensified charge coupled device. The blue (red) path represents the 405 nm (810 nm) photons. (b) Coincidence interference curves for entanglement. (c) and (d) Cross-polarization image and scanning electron microscope (SEM) image of metasurface M$_B$ ($q = 1/2$) (Scale bar, 500 nm). (e) and (f) Theoretical and experimental optical axes distribution image of M$_B$.

To verify the principle, our experimental setup is depicted in Fig.2(a). It comprises three stages: entangled photon source, polarization remote trigger, and generation and measurement of diverse quantum path entanglement. The entangled photon source can be generated from SPDC process in a Sagnac configuration, where the type-II down-conversion periodically poled KTiOPO$_4$ (PPKTP) crystal is pumped by a 405 nm single-frequency diode laser. To maintain balance in the pump power between clock-wise and counterclockwise direction, the combination of a quarter wave plate (QWP) and a half wave plate (HWP) can be adjusted in front of the Sagnac loop. The Sagnac interferometer loop consists of two broadband dielectric mirrors (BDM), a dual-wavelength polarization beam splitter (DPBS), and a dual-wavelength half-wave plate (DHWP). By fixing the DHWP at 45°, the horizontally polarized pump light is obtained in the counterclockwise direction. The entangled photon pairs are spatially separated by DPBS, with the signal and idler photons as outputs. In polarization remote trigger section: idler photon of entangled photon pair is continuously collapsed into any paths for remote control, and the other, signal photon is coupled to single-mode fiber, delayed by 17 m and brought to the system for generation and measurement of diverse quantum path entanglement. Here, the entangled photon pair state $|\Psi\rangle=1/\sqrt{2}(|H\rangle_i|H\rangle_s+|V\rangle_i|V\rangle_s)$ is achieved by adjusting the combination of HWP and QWP in signal arm.

To characterize the quality of the entangled state, we performed two-photon polarization interference. In the coincidence counts, the system for generation and measurement of diverse quantum path entanglement is replaced by a single photon avalanche detector (SPAD) (see details in Supplementary Material). The coincidence interference curves can be measured for entanglement, as shown in Fig.2(b) (see measured details in Supplementary Material). The visibilities of the interference fringes are calculated using the formula $V = (C_{max} - C_{min})/(C_{max} + C_{min})$, where $C_{max}$ and $C_{min}$ represent maximum and minimum coincidence counts, respectively. The visibilities of interference fringe are $V_{D/A} = 96.8 \pm 0.1\%$ and $V_{H/V} = 99.8 \pm 0.1\%$, respectively (see calculated details in Supplementary Material). When the visibilities exceed 71%, the bound required to violate Bell's inequality is met [3,55]. It is proved that the quality of the generation of entangled state is pretty high and stable.

Noncommutative metasurfaces are utilized to create diverse quantum path entanglement, due to its multifunctionality and non-commutativity. As shown in Fig.2(c), the local optical slow-axis orientation of metasurface $M_B$ ($q = 1/2$) is proved by using the polariscope method. In addition, the transverse gradient pattern of the optical axis is emphasized by crossed linear polarizer imaging under $100 \times$ magnification. The nanostructures are roughly in the range of $30-100$ $nm$, and Fig.2(d) provides the scanning electron microscope (SEM) image of metasurface $M_B$. To fully characterize the structure of metasurface $M_B$, its optical axis distribution is measured experimentally as shown in Fig.2(e), which has a good agreement with the theoretical results [Fig.2(f)].

## The generation of quantum path entanglement of structured photons based on noncommutative metasurfaces

To experimentally demonstrate the ability to generate quantum path entanglement of structured photons by metasurfaces, we consider combination of two metasurfaces, $M_B$ followed by $M_A$ as shown in Fig.1(f), and placed in signal arm in Fig.2(a). As an example, the idler photons are continuously collapsed into a cardioid path on fundamental PS by adjusting QWP and P in trigger remote control section, as illustrated on the left of Fig. 3. According to Eq. 8, because of quantum path entanglement, structured signal photons and the idler photons along the entangled path are mutually entangled. The corresponding cardioid path on HOPS with order $m = -1$ can be nonlocally measured [on the right of Fig. 3], through the collapse of the idler photon's state in continuous moment. The ability to mutually entangle between signal photons and idler photons has been illustrated by coincidence interference curve in Fig.2(b). Here, consider that idler photons are collapsed into eight specific polarization states along a cardioid path on fundamental PS, with their polarization state shown in Figs.3(a)-3(h). Because of the quantum path entanglement between idler photons and structured signal photons, the corresponding structured signal photons state of eight specific polarization along a cardioid path can mirror idler photons on HOPS with order $m = -1$. Figures 3(a1)-3(h1) provide the corresponding experimental polarization patterns of structured signal photons states, which have good agreement with theoretical results (see details in Supplementary Material). As shown in Fig.2(a), QWP, P, and ICCD in signal arm is used to perform Stokes parameters measurement of structured signal photons, which is utilized to reconstruct its complex polarization pattern (see the details and principles of reconstruction in Supplementary Material). It is confirmed that quantum path entanglement can be realized by the interaction between noncommutative metasurfaces and signal photons. In addition, the shape of quantum path entanglement can be nonlocally changed and measured by idler photons remotely controlling.

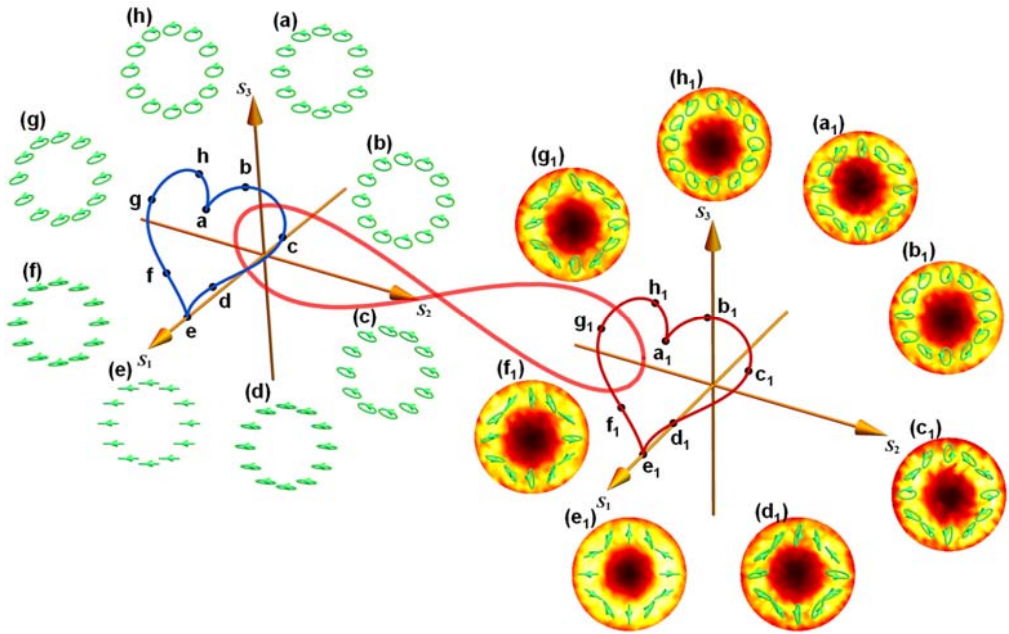

**Fig.3. Generation and measurement of quantum path entanglement of structured photons based on metasurfaces.** (a)-(h) The collapsed polarization state distribution of idler photons at any eight points along cardioid path of fundamental PS. (a1)-(h1) The corresponding reconstructed polarization patterns and intensity distribution of structured signal photons on HOPS with order $m = -1$. Cardioid path is mutually entangled between idler photons and structured signal photons.

## Noncommutative metasurfaces enabled diverse quantum path entanglement of structured photons

It has been experimentally demonstrated that metasurfaces can realize quantum path entanglement of structured photons. Building on this, our investigation is extended to the tunability of diverse quantum path entanglement across different HOPS by exploiting non-commutativity of metasurfaces. Specifically, we consider metasurafce $M_A$ with uniform optical axis distribution and metasurface $M_C$ designed as a q-plate with $2\pi$ phase retardation, as depicted in top of Fig.4. When the $M_A$ is placed before $M_C$ in signal arm [top left inset of Fig.4], diverse quantum path entanglement can be generated between fundamental PS and HOPS with order $m = 2$. Because of diverse quantum path entanglement between them, the structured signal photons can be nonlocally controlled and real-time manipulated to correspond to any states on the HOPS as the idler photons continuously collapses into state on fundamental PS. As an example, equator as a special evolvement path is mutually entangled between idler photons and structured signal photons. Figures 4(a)-4(d) show the simulated polarization pattern results for structured signal photons at four certain points along equator of HOPS with order $m = 2$, triggered by the corresponding polarization state of the idler photons as indicated by the white arrow. The complex polarization pattern of the structured signal photons is reconstructed using Stokes parameters, with experimental results shown in Figs. 4(e)-4(h). **Fig.4.**

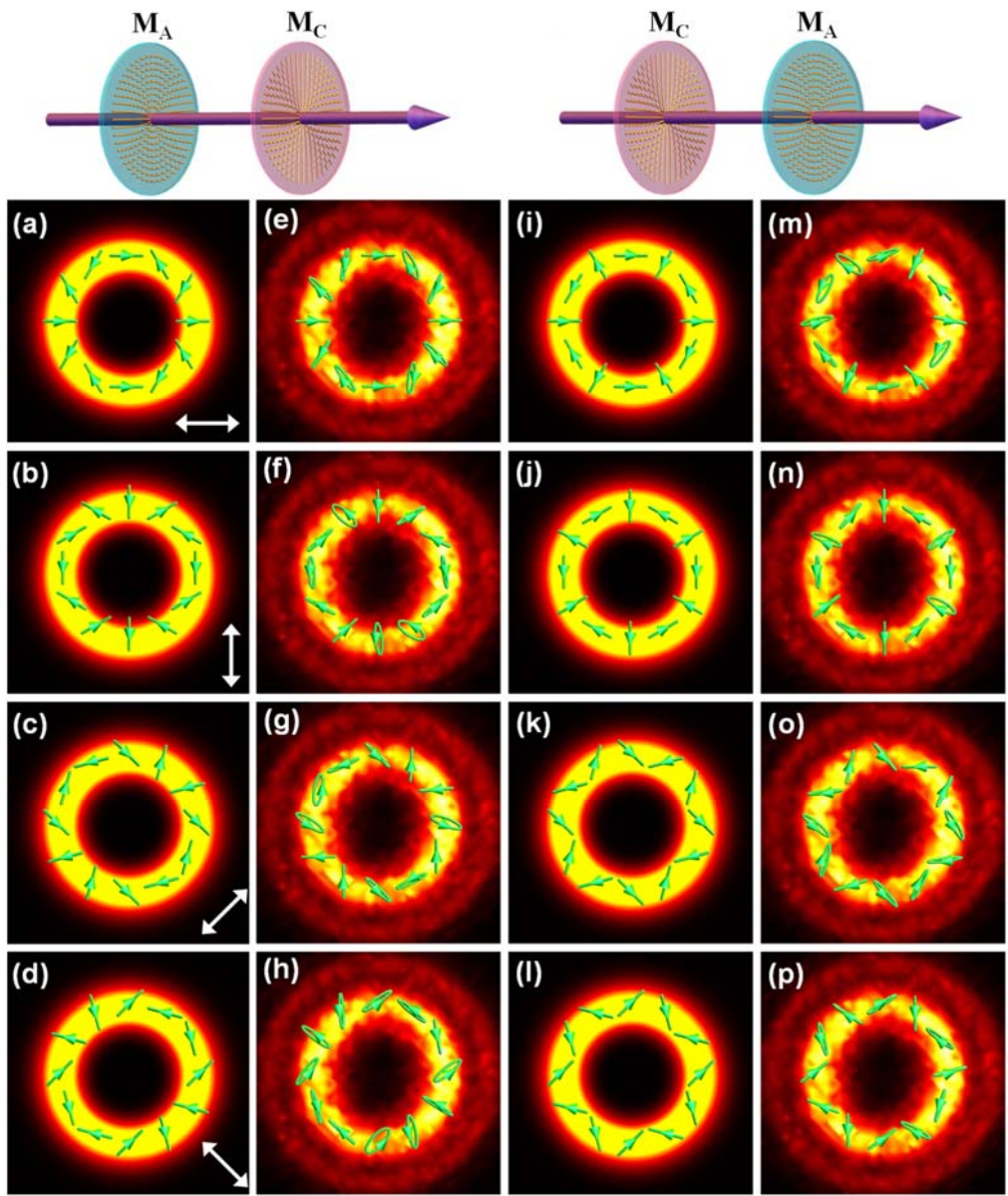

**Experimental demonstration for diverse quantum path entanglement of structured photons across HOPS with opposite order via noncommutative metasurfaces.** Top inset: schematic representations of the two distinct sequences of cascaded metasurfaces: $M_A M_C$ (top left), $M_C M_A$ (top right). These sequences indicate the noncommutative property of metasurfaces, which enables the diverse switching of

quantum path entanglement across HOPS with opposite orders. (a)-(d) and (e)-(h) Theoretical and experimental polarization maps for the structured signal photons at four specific points along the equator of HOPS with order $m = 2$, corresponding to the sequence $M_A M_C$ (i)-(l) and (m)-(p) Theoretical and experimental polarization maps for the structured signal photons at four specific points along the equator of HOPS with order $m = -2$, corresponding to the sequence $M_C M_A$. White arrow direction represents the polarization states of the idler photons that trigger the corresponding structured signal photons polarization patterns.

When the sequence of cascaded metasurfaces $M_A$ and $M_C$ is reversed (top right inset of Fig.4), their noncommutativity switches the diverse quantum path entanglement from HOPS with order $m = 2$ to $m = -2$ as illustrated in Figs. 4(i)-4(l). The corresponding experimental results as shown in Figs. 4(m)-4(p), are in good agreement with the simulated results. The results confirm that the noncommutativity of cascaded metasurfaces can realize and switch diverse quantum path entanglement of structured photons between HOPS of opposite orders.

Building upon our demonstration of diverse quantum path entanglement across HOPS with opposite orders, the versatility of diverse quantum path entanglement can be extended across a broader spectrum of HOPS. Here, three metasurfaces are introduced to further exploit the noncommutative property, enabling the switching of diverse quantum path entanglement not only between opposite orders but also across distinct non-opposite orders of HOPS. Three special combination order of cascaded metasurfaces is considered, including $M_A M_B M_C$, $M_B M_A M_C$, $M_B M_C M_A$, as shown in top inset of Fig.5. Compared with the previous two metasurfaces cascaded combination, $M_B$, designed as a q-plate with π phase retardation, is added. Due to metasurfaces with noncommutativity, diverse quantum path entanglement across different orders of HOPS can be created and tunable by changing the combination order of three cascaded metasurfaces (theoretically derived details see Supplementary Material).

When the combination order of cascaded metasurfaces is $M_A M_B M_C$ (top left inset of Fig.5), diverse quantum path entanglement can be realized on the HOPS with order $m = 1$. Figures 5(a)-5(d) and 5(e)-5(h) display theoretical and experimental results of maps and intensity distribution for the structured signal photons at four specific points along the equator of HOPS with order $m = 1$. They can be triggered by idler photons, and white arrows correspond to different collapsed polarization state of the idler photons. Analogously, for the combination order of cascaded metaurfaces $M_B M_A M_C$ (top center inset of Fig.5), diverse quantum path entanglement can be transformed onto the HOPS with order $m = 3$, with the theoretical and experimental results shown in Figs.5(i)-5(l) and 5(m)-5(p). The combination sequence of cascaded metaurfaces, $M_B M_C M_A$ (top right inset of Fig.5), can be utilized to create diverse quantum path entanglement on the HOPS with order $m = -1$, with the theoretical and experimental results presented in Fig.5(q)-5(t) and 5(u)-5(x). All reconstructed polarization distributions of structured signal photons reveal a good agreement with the theory, which proves the versatility of diverse quantum path entanglement manipulation across a broader order of HOPS. This extension facilitates the control and manipulation of quantum states, offering unprecedented flexibility in the generation of diverse entangled states with multiple DOFs.

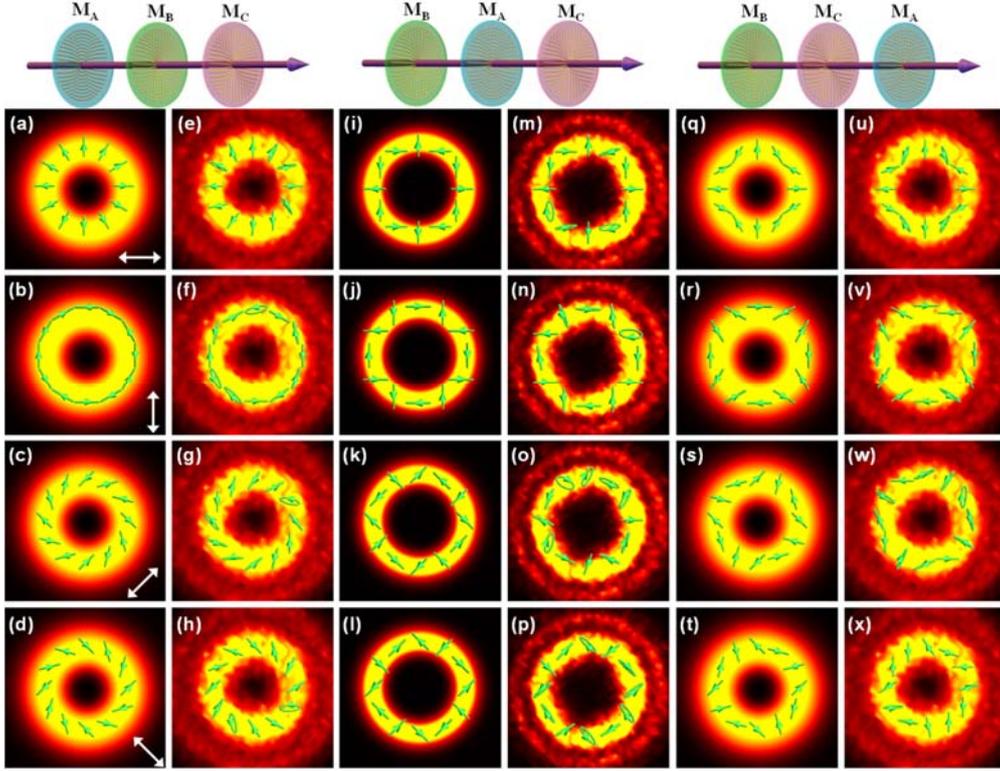

**Fig.5. Generation of diverse quantum path entanglement of structured photons across multiple HOPS via noncommutative metasurfaces.** Top Inset: schematic representations of the three distinct sequences of cascaded metasurfaces: $M_A M_B M_C$ (top left), $M_B M_A M_C$ (top center), $M_B M_C M_A$ (top right). These sequences highlight the noncommutative property of metasurfaces, which enables the switching of diverse quantum path entanglement across different HOPS orders. (a)-(d) and (e)-(h) Theoretical and experimental polarization maps for the structured signal photons at four specific points along the equator of HOPS with order $m = 1$, corresponding to the sequence $M_A M_B M_C$. (i)-(l) and (m)-(p) Theoretical and experimental polarization maps for the structured signal photons at four specific points along the equator of HOPS with order $m = 3$, corresponding to the sequence $M_B M_A M_C$. (q)-(t) and (u)-(x) Theoretical and experimental polarization maps for the structured signal photons at four specific points along the equator of HOPS with order $m = -1$, corresponding to the sequence $M_B M_C M_A$. White arrow direction represents the polarization states of the idler photons that trigger the corresponding structured signal photons polarization patterns.

**Discussion**

In conclusion, we have theoretically and experimentally demonstrated the generation for diverse quantum path entanglement of structured photons based on noncommutative property of metasurfaces. By exploiting the interaction of metasurfaces with entangled photons, any shape of diverse quantum entanglement path has been experimentally generated between fundamental PS and different HOPS. This capability is pivotal for the advancement of quantum communication and computation, particularly control and manipulation of entangled states. The reconstructed polarization distributions for the structured signal photons at specific points along entangled path of HOPS showcase the high precision and feasibility of our technique. This precision is essential for the reliable generation of entangled states, which are the building blocks of quantum technologies.

Benefiting from noncommutative capacity of metasurfaces, diverse quantum path entanglement has been demonstrated across different HOPS orders, opening up new avenues for the implementation of complex quantum protocols. Our findings suggest that noncommutative metasurfaces can enable the creation of diverse path entanglement, a key component in the development of next-generation quantum devices. Furthermore, the ability to control diverse quantum path entanglement across a broader range of HOPS orders could lead to the exploration of more complex quantum operations and the realization of more sophisticated quantum computing architectures. In briefly, our work underscores the potential of noncommutative metasurfaces in the manipulation of diverse entangled states.

**Funding.** National Natural Science Foundation of China (Grant No. 12174097).

**Disclosures.** The authors declare no conflicts of interest.

**Data availability.** Data underlying the results presented in this paper are not publicly available at this time but may be obtained from the authors upon reasonable request.

**Supplementary Document.** See Supplementary Material for supporting content.